\documentclass[conference]{IEEEtran}
\IEEEoverridecommandlockouts
\usepackage{cite}
\usepackage{amsmath,amssymb,amsfonts}
\usepackage{algorithm}
\usepackage{algorithmic}
\usepackage{graphicx}
\usepackage{textcomp}
\usepackage{xcolor}
\usepackage{mathtools}
\usepackage{dsfont}
\usepackage{multicol}
\usepackage{caption}
\usepackage{comment}
\usepackage{subcaption}

\def\BibTeX{{\rm B\kern-.05em{\sc i\kern-.025em b}\kern-.08em
    T\kern-.1667em\lower.7ex\hbox{E}\kern-.125emX}}


\begin{document}
\bstctlcite{IEEEexample:BSTcontrol}

\title{Learning-Based Latency-Constrained Fronthaul Compression Optimization in C-RAN\\
\thanks{This work has received funding from the European Union's Horizon Europe research and innovation programme under the Marie Skłodowska-Curie grant agreement No. 101073265.}
}

\author{\IEEEauthorblockN{Axel Gr\"onland\textsuperscript{1,2}, Bleron Klaiqi\textsuperscript{2}, Xavier Gelabert\textsuperscript{2}}
\IEEEauthorblockA{\textit{\textsuperscript{1}Royal Institute of Technology (KTH), Stockholm, Sweden} \\ \textit{\textsuperscript{2}Huawei Technologies Sweden AB, Stockholm Research Centre, Sweden} \\
gronland@kth.se, \{bleron.klaiqi, xavier.gelabert\}@huawei.com}
}

\maketitle

\begin{abstract}
The evolution of wireless mobile networks towards cloudification, where Radio Access Network (RAN) functions can be hosted at either a central or distributed locations, offers many benefits like low cost deployment, higher capacity, and improved hardware utilization. Nevertheless, the flexibility in the functional deployment comes at the cost of stringent fronthaul (FH) capacity and latency requirements. One possible approach to deal with these rigorous constraints is to use FH compression techniques. 
To ensure that FH capacity and latency requirements are met, more FH compression is applied during high load, while less compression is applied during medium and low load to improve FH utilization and air interface performance. 
In this paper, a model-free deep reinforcement learning (DRL) based FH compression (DRL-FC) framework is proposed that dynamically controls FH compression through various configuration parameters such as modulation order, precoder granularity, and precoder weight quantization that affect both FH load and air interface performance. Simulation results show that DRL-FC exhibits significantly higher FH utilization (68.7\% on average) and air interface throughput than a reference scheme (i.e. with no applied compression) across different FH load levels. At the same time, the proposed DRL-FC framework is able to meet the predefined FH latency constraints (in our case set to 260 $\mu$s) under various FH loads.
\end{abstract}

\begin{IEEEkeywords}
C-RAN, fronthaul, machine learning, reinforcement learning, compression, performance evaluation.
\end{IEEEkeywords}

\section{Introduction}
\label{sec:introduction}
Centralized Radio Access Network (C-RAN) deployments \cite{CRAN_ChinaMobile} offer substantial cost savings by allowing the dissagregation of RAN  functionalities where the processing of such functionalities is \emph{split} between the remote radio units (RRU), close to the antenna masts, and the baseband unit (BBU), at some centralized location. Then, a centralized pool of BBUs can exploit statistical multiplexing gains by jointly processing RAN functions from a large number of RRUs, under the assumption that not all RRUs will be subject to high-load conditions simultaneously. This allows a better resource utilization and dimensioning than non-centralized (a.k.a. distributed) RAN options, where all processing is done locally, and where each BBU should be dimensioned for peak requirements of each RRU. In addition, C-RAN offers, among other features, increased maintainability, flexibility and upgradability, as well as improved and fast coordination features such as CoMP, inter-cell handover, etc. \cite{CRAN_ChinaMobile,checko_cloud_2015}. 
In contrast, C-RAN may cause high data rate requirements on the fronthaul (FH), which interfaces between the RRU and the BBU, as well as increased latency budget constraints in the overall signal processing chain \cite{checko_cloud_2015}. Indeed, a major challenge in C-RAN deployments is the huge demands on bandwidth aggregation required for the FH, especially for specific split options. As noted in \cite{duan_performance_2016}, for a fully centralized split in a 5G New Radio (NR) system, with 100 MHz bandwidth and 64 antennas, the required FH transmission rate exceeds several hundreds of Gbps, for a single cell. A thorough characterisation,
modelling and evaluation of functional splits in a flexible RAN for dynamic functional split optimisation in 5G and beyond systems is presented in \cite{func_split_tutorial_2023}.
A site serving several cells would require a yet increased aggregated FH data rate. For this, more favourable splits can be defined \cite{rodriguez_cloud-ran_2020}, combined with data compression methods \cite{lorca_lossless_2013, lagen_fronthaul-aware_2021,lagen_fronthaul_2022} that can help alleviate the data rate demands over the FH link.

Current literature has tackled the aforementioned challenges in several ways. In \cite{liu_graph-based_2015}, a graph-based framework is proposed to effectively reduce the FH cost through properly splitting and placing baseband processing functions in the network.
In \cite{lorca_lossless_2013}, a lossless compression technique for the FH is presented, which depends on the ratio of occupied resources. Authors in \cite{lagen_modulation_2021} provide insights on FH compression based on so-called modulation compression whereby the modulation constellations are conveniently encoded so as to reduce the the required FH capacity, reportedly up to 82\%. Work in \cite{lagen_fronthaul_2022} combines modulation compression with scheduling strategies to further optimize the use of FH-limited deployments. Joint FH compression and precoding design is proposed in\cite{Kang_FH_Comp_Prec_2016}, where two different splits, determining the location of the precoder, are investigated. Noteworthy, the previous works are based on conventional mathematical optimization approaches that, in general, exhibit very high complexity to obtain optimal solutions and heavily rely on the availability of underlying models, which are difficult to acquire for realistic scenarios.

Accordingly, we resort to Machine Learning (ML) approaches to tackle the FH compression optimization problem. The use of ML techniques to address complex optimization problems in wireless networks has been widely covered in the literature, see e.g. \cite{ali_6g_2020} and references therein. Notably, regarding the C-RAN domain, in \cite{Matoussi_dl_ue_slice_2020}, UE slicing and functional split optimization has been investigated using supervised learning, though requiring high quality labels which are difficult and costly to obtain in practice. A framework based on model-free reinforcement learning (RL) to jointly select the most suitable split and computing resource allocation is proposed in \cite{murti_learning-based_2022}.

In this paper, we will address the configuration and subsequent optimization of the FH link in the DL of a C-RAN deployment. We propose model-free deep RL-based FH compression (DRL-FC) framework as a data-driven solution framework that dynamically adjusts several parameters related to compression schemes, including modulation order, precoder granularity, and precoder weight quantization. All of this, with the objective of maximizing FH utilization and herewith maximize air interface throughput while keeping FH latency below some predefined target (in our case 260 $\mu$s). Compared to previous work, to the best of our knowledge, our contribution is novel in that it addresses the problem of FH compression optimization with learning-based methods while at the same time targeting at guaranteeing latency constraints.

The remainder of this paper is organized as follows. In Section \ref{sec:system_models} we provide a description of the assumed C-RAN scenario, outline our simulation setup and describe our system models. The optimization problem formulation is described in Section \ref{sec:prob_formulation}. Herein we derive a RL problem formulation and show that it is equivalent. Subsequently, we derive a reward function, mapping it to the problem we are trying to solve, as well as some additional techniques we use in order to stabilize and speed-up learning. Hereafter, in Section \ref{sec:numerical_evaluation}, we provide results of our experiments with an emphasis on how much throughput gain we could expect compared to other more static policies. Lastly, in Section \ref{sec:conclusions} we outline some concluding remarks.

\section{Scenario and System Model}
\label{sec:system_models}
\subsection{C-RAN scenario description}\label{sec:Scenario_Desc}
We consider the DL direction in a Time Division Duplex (TDD) system following the 3GPP NR standard. In the frequency domain, a chosen subcarrier spacing (SCS), defined as $\Delta f_{\textsc{scs}}=15\cdot2^\mu$ (kHz) with SCS index $\mu=\{0,1,2,3,4\}$, partitions the available bandwidth $B$ (Hz) into a number of $N^{\textsc{prb}}_{\textsc{B},\mu}$ Physical Resource Blocks (PRBs). Each PRB contains 12 subcarriers. In the time domain, the TDD frame-structure periodically cycles over time slots, carrying DL and UL data, each with a duration of 14 symbols, i.e. $T_{slot}^\mu=14\cdot T_{symb}^\mu $, with $T_{symb}^\mu$ the duration of a symbol. NR also defines the number of subcarrier-symbol pairs  (a.k.a. Resource Elements, REs) contained in a single PRB during the duration of a slot, thus $N_{\textsc{re}}=12\cdot14=168$. Multi-User MIMO (MU-MIMO) with $N_{ant}$ at the transmitter is supported via digital beamforming by precoding the user data with pre-calculated weights based on channel estimations from UL pilot measurements \cite{khorsandmanesh_quantization-aware_2022}. This allows up to $\upsilon_{lay}\le N_{ant}$ spatially multiplexed users to be scheduled at the same time, over the same frequency resource.

\begin{figure}
    \centering
    \includegraphics[scale=0.9]{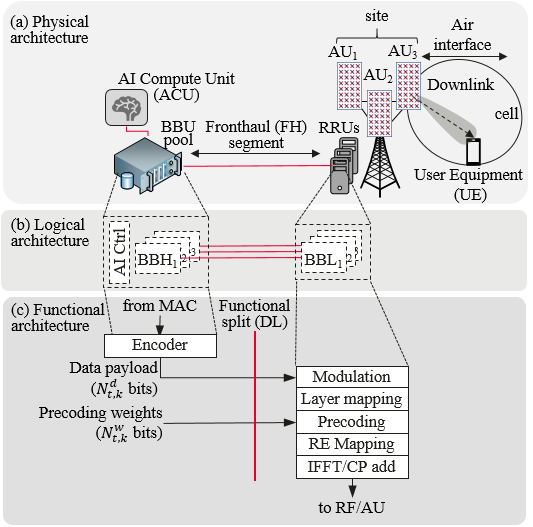}
    \caption{The considered scenario with the (a) physical architecture; (b) logical architecture and (c) functional architecture.}
    \label{fig:scenario}
\end{figure}

As mentioned in Sec. \ref{sec:introduction}, we draw our attention towards the C-RAN architecture \cite{CRAN_ChinaMobile}, where the baseband processing for a number of $K$ geographically distributed RRUs (each serving a cell) can be \emph{split} between a centralized location, where a pool of BBUs reside, and each of the said RRUs.  Fig. \ref{fig:scenario}(a) depicts the physical architecture for a simple $K=3$ cell case scenario, where the RRU and BBU are interconnected via the FH. The FH, characterized by an available FH capacity of $C_{\textsc{fh}}$ (Gb/s), carries the aggregate DL generated traffic towards all three cells in this example. Accordingly, the term \emph{split} will be used hereafter to describe the amount of baseband processing functions residing in each of the aforementioned locations. The FH may consist of a packet-oriented networked system, including switches and links following a specific networking protocol, e.g. eCPRI \cite{duan_performance_2016}. From a logical architecture perspective, see Fig. \ref{fig:scenario}(b), and for each cell served by a RRU, we can consider the Baseband Low (BBL) entity, residing at each RRU, and the Baseband High (BBH) entity residing at the pool of BBUs. 
Similar to \cite{lorca_lossless_2013, khorsandmanesh_quantization-aware_2022}, in this work we consider a DL split where the encoded user data and the precoding weights are sent separately over the FH (see Fig. \ref{fig:scenario}(c)). The encoded user data (in bits) can be mapped to the corresponding modulated symbols at the BBL, thus experiencing no quantization errors over the FH. The precoding weights however, which are complex-valued, need to be quantized with some bit resolution $b^w$, and are therefore prone to quantization errors. 
\subsection{FH compression configuration}\label{sec:Compression_config}
The FH is usually dimensioned taking into account the statistical multiplexing gain due to the spatial traffic randomness at different cells in a given area \cite{wangFronthaulStatisticalMultiplexing2017}. In practice, this means that the FH is underdimensioned, i.e., it cannot uphold the transmission of peak data rates towards all fully-loaded cells simultaneously. Then, in order to keep the FH throughput below the dimensioned capacity, compression methods can be, and have been, introduced. In particular, three of such methods will be considered in this paper. Firstly, the modulation order, which in NR is given by the set $\mathcal{Q}^{\textsc{NR}}=\{2,4,6,8\}$ bit/symbol \cite{3gpp_ts_38.211_v15.5.0_nr}, can be restricted to some maximum value $q\in{\mathcal{Q}^{\textsc{NR}}}$ so that symbols are represented with fewer bits, thus lowering the FH throughput. This will, of course, have an impact on the air interface throughput, but its use can be limited to specific time instants where the FH cannot support the offered traffic. Second, we can modify the bitwidth $b^w\in\mathbb{N}$ (expressed in bits) for precoding weights used to quantify and represent complex-valued samples for their transmission over the FH. 
Finally, one can tune the sub-band precoding granularity, $r^w$, which reflects the number of subcarriers that can be assumed to share similar channel conditions and therefore admit the same precoding weight value \cite{lorca_lossless_2013}. We define the sub-band precoding granularity as the number of consecutive PRBs being applied the same precoding weight. For example, $r^w=1$ indicates that each PRB in the entire bandwidth will be precoded with its own distinctive weight, whereas $r^w=2$ means that two consecutive PRBs will be precoded with the same weight (effectively halving the weight payload to be transmitted over the FH), and so on.

Given FH compression will encompass some degree of air interface performance degradation, it is important to only make use of it when the situation requires it, i.e., when the FH utilization (the ratio between used and available FH capacity) is high and there is risk of not being able to deliver packets in time over the FH. At medium and low FH loads, compression should be reduced leading to increased, yet below limit, FH utilization and consequently better air interface performance. Maximisation of the FH utilization constrained on FH latency is therefore targeted as the main optimization criteria hereon.

\subsection{FH utilization}
With the assumed functional split given in Sec. \ref{sec:Scenario_Desc}, the FH will carry the aggregated traffic towards the different cells, which comprises data payload and precoding weights. The data payload (in bits) to be delivered at a given time slot $t$ intended for the cell $k$ is given by:
\begin{align}
N_{t,k}^d = N_{\textsc{re}}\cdot\upsilon_{lay}\cdot N_{t,k}^{\textsc{prb}}\cdot q_{t,k},
\end{align}
where $\upsilon_{lay}$, $N_{t,k}^{PRB}$, and $q_{t,k}$ denote number of layers, number of the allocated PRBs for a given slot $t$ and cell $k$, and modulation order for slot $t$ and cell $k$, respectively.

In a similar way, the number of precoding weight bits transmitted at slot $t$ towards cell $k$ can be obtained as follows:
\begin{align}
N_{t,k}^w = \Biggl\lceil\frac{N_{t,k}^{\textsc{prb}}}{r_{t,k}^w}\Biggr\rceil\cdot \upsilon_{lay}\cdot N_{ant}\cdot b_{t,k}^w,
\end{align}
with $r_{t,k}^w$, $N_{ant}$, and $b_{t,k}^w$, the precoder granularity, the number of antennas, and the weight bit quantization, respectively.

The FH data rate at slot $t$ for cell $k$ is given by:
\begin{align}
R_{t,k}^{\textsc{fh}}= \frac{1}{T_{slot}^\mu}\left(N_{t,k}^d+N_{t,k}^w\right),
\end{align}
where $T_{slot}^\mu$ is the slot duration.

Finally, the average FH utilization over $T$ slots and $K$ cells can be calculated as follows:
\begin{equation}
\label{eq:avg_util}
\overline{\rho} = \frac{1}{T}\sum\limits_{t=1}^{T}\sum\limits_{k=1}^{K}\rho_{t,k}^{\textsc{fh}} = \frac{1}{T}\sum\limits_{t=1}^{T}\sum\limits_{k=1}^{K}\frac{R_{t,k}^{\textsc{fh}}}{C_{\textsc{fh}}},
\end{equation}
with $\rho_{t,k}^{\textsc{fh}}$ the FH utilization for slot $t$ and cell $k$. $C_{\textsc{fh}}$ and $K$ denote the FH capacity and the number of cells sharing the FH, respectively.

\section{Optimization Formulation and Proposed DRL-FC Framework}
\label{sec:prob_formulation}
Besides the functional split, other parameters might affect the system performance and FH load. Therefore, finding the optimal configuration of the system that maximizes the average FH utilization under a FH latency constraint is a difficult task, more so in a highly dynamic environment. For the DL direction, FH latency is defined as the time elapsed between a packet being sent by the BBH and its reception at the corresponding BBL. As mentioned earlier, in this paper we focus on the DL transmission and aim to maximize the average DL FH utilization while keeping FH latency under a certain threshold.

Let $c_{t,k} = (q_{t,k},b_{t,k}^w,r_{t,k}^w)$ denote the compression configuration and $L_{t,k}$ the FH latency at slot $t$ for cell $k$. We denote $\mathcal{\tau}=\max\limits_{t,k}L_{t,k}$ as the maximum FH latency observed over $T$ slots for $K$ cells. The problem formulation can be written as the following constrained optimization problem:
\begin{align}
&\max_{c_{1,1},\ldots,c_{\textsc{T},\textsc{K}}}\overline{\rho},\label{eq:Objective}\\
&\textrm{s.t.}\nonumber\\ 
&\mathbb{P}\big(\tau>\tau_{max}\big) \leq \delta,\label{eq:Constrain_1}\\
&\sum\limits_{k=1}^{K}\rho_{t,k}^{\textsc{fh}}\leq 1, q_{t,k}\in \mathcal{Q}, b_{t,k}^w\in\mathcal{B}^w, r_{t,k}^w\in \mathcal{R}^w.\label{eq:Constrain_4}
\end{align}
Here, $\tau_{max}$ denotes the maximum allowed FH latency, and $\mathcal{Q}$, $\mathcal{B}^w$, along with $\mathcal{R}^w$ are the sets of used modulation orders, weight bits, and precoder granularities, respectively. The first constraint \eqref{eq:Constrain_1} ensures that the probability of FH latency surpassing maximum allowed latency ($\tau_{max}$) during $T$ slots and $K$ cells does not exceed a specified target $\delta$.

In general, finding the optimal solution for the constrained optimization problem \eqref{eq:Objective}-\eqref{eq:Constrain_4} is computationally challenging. Furthermore, it is very difficult to obtain an accurate close-form expression for the FH latency due to dependency on many parameters such as functional split, FH topology, switch implementation etc. Then, a numerical evaluation of the FH latency, i.e., $L_{t,k}$, will be obtained through our in-house simulator, as will be explained in Section \ref{sec:numerical_evaluation}. 

We aim to solve the constrained optimization problem \eqref{eq:Objective}-\eqref{eq:Constrain_4} via a model-free RL scheme, which treats the underlying system as a \emph{black-box} and requires neither a system model nor collection and labeling of data, unlike supervised learning models. Next, details of the proposed scheme are presented.

\subsection{The proposed DRL-FC framework}
Recalling from Sec. \ref{sec:prob_formulation}, the compression configuration is controlled via three different parameters: $q_{t,k}$, $b_{t,k}^w$, and $r_{t,k}^w$. Then, the system state can be defined as: 
\begin{align}
s_t=\{\rho_{t, k}^{\textsc{fh}},L_{t,k}, q_{t,k}, b_{t,k}^w, r_{t,k}^w | k=1,2,3...K\},
\end{align}
which comprises the FH utilization, the FH latency and the configuration at slot $t$ and cell $k$. The action at time $t$ is defined as a change in configuration as follows:
\begin{align}
a_t = \{\Delta q_{t,k}, \Delta b_{t,k}^w, \Delta r_{t,k}^w \in \{-1, 0, 1\}| k=1,2,...K\}.
\end{align}
Here, $\Delta q_{t,k}$, $\Delta b_{t,k}^w$, and $\Delta r_{t,k}^w$ denote changes in parameters $q_{t, k}$, $b_{t,k}^w$, and $r_{t,k}^w$, respectively. More specifically, $\Delta q_{t,k} = -1$ causes a change to a lower modulation order, $\Delta q_{t,k} = 1$ means moving to higher modulation order while $\Delta q_{t,k} = 0$ implies no change. Similar applies to $\Delta b_{t,k}^w$ and $\Delta r_{t,k}^w$.

Let $|\mathcal{Q}|$, $|\mathcal{B}^w|$, and $|\mathcal{R}^w|$ denote cardinality of the set of values for $\mathcal{Q}$, $\mathcal{B}^w$, and $\mathcal{R}^w$, respectively. The cardinality of action space $\mathcal{A}$ for $K$ cells is then given by:
\begin{align}
|\mathcal{A}| = \left(|\mathcal{Q}|\times |\mathcal{B}^w|\times |\mathcal{R}^w|\right)^K.
\end{align}
In view of this, the action space size can get very large, even for a moderate number of parameter values. For example, for $|\mathcal{Q}|=2$, $|\mathcal{B}^w|=7$ and $|\mathcal{R}^w|=3$, the total number of possible actions is $|\mathcal{A}|=42^K$. 
Assuming all cells experience similar load and that we change one parameter at a time for each cell, the number of possible actions is reduced to $|\mathcal{\Tilde{A}}|=7$, which is reasonable for efficient Q-learning implementation. 

Next, we reformulate the average horizon problem of (\ref{eq:avg_util})-(\ref{eq:Objective}) into an equivalent discounted infinite horizon problem. In order to incorporate the above constraints, the Lagrangian can be applied to formulate an optimization problem \cite{boyd2004convex}. We can express the Lagrangian via a discounted sum over a reward function $ r_\lambda(s_t, a_t)$ as:
\begin{equation}
    \label{eq:lagrangian} 
    \min_{\lambda > 0}\max_\pi\mathcal{L}(\pi, \lambda) =\mathbb{E}\Big[\sum_{t=0}^\infty \gamma^t r_\lambda(s_t, a_t)\Big|\pi\Big],
\end{equation}
where
\begin{align}
    r_\lambda(s_t, a_t) &=\sum_{k=1}^{K}\rho^{\textsc{fh}}_{t, k} + \lambda g(s_t),\\
    g(s_t) &= \mathds{1}(\tau < \tau_{max}) - d.
\end{align}
We choose $d$ in accordance with\cite{safeRL} to ensure the optimal policy $\pi^*$ does not violate the constraint \eqref{eq:Constrain_1}. Discounted infinite horizon problems are known to be equivalent to the average reward problems with $\gamma$ sufficiently close to 1 \cite{kakade2001avg_reward}.

The Lagrangian in (\ref{eq:lagrangian}) can be solved with RL techniques through alternating between optimizing over the policy and the Lagrange multipliers\cite{boyd2004convex}. Due to the discrete action space, we use value-based methods which are more sample-efficient than policy gradient methods\cite{haarnoja2018soft, sutton2018reinforcement}. We also set the Lagrange multipliers to 1, as FH utilization for $K$ cells at slot $t$ is upper bounded by 1, see \eqref{eq:Constrain_4}. We do this to avoid having to optimize the Lagrange multiplier in parallel with the policy, which we found in our experiments to be very challenging. In DQN we define our training objective as:
\begin{equation}
\label{eq:JQ}
    J_Q = \mathbb{E}_{(s_i, a_i, s'_i) \sim \mathcal{D}} \Big[\frac{1}{2}\big(y_{\Bar{\theta}}(s_i, a_i, s'_i) - Q_\theta(s_i, a_i)\big)^2\Big],
\end{equation}
where $y_{\Bar{\theta}}$ is the Bellman backup\cite{sutton2018reinforcement} which in this work is modified to be the 
Double Deep Q-Network (DDQN) \cite{van2016deep} targets. We also utilize \emph{prioritized experience replay} \cite{schaul2015prioritized} and \emph{Boltzmann exploration} which is the behaviour policy given by:
\begin{equation}
    \label{boltzman_explore}
    \beta_\alpha(a|s_t) = \frac{e^{Q(s_t,a)/\alpha}}{\sum_{b\in \mathcal{A}} e^{Q(s_t,b)/\alpha}}.
\end{equation}
The constant $\alpha$ is the temperature which controls the entropy of the distribution and is annealed towards zero during the training. The distribution will tend towards the greedy policy for $\alpha \rightarrow 0$. In our experiments these techniques had the most impact on the learning speed/performance of the algorithm and are therefore highlighted in this work. The training procedure is summarized in Algorithm \ref{alg:DDQN}. 
\begin{algorithm}
\caption{DRL-FC based on DDQN with prioritized experience replay and Boltzmann exploration.}\label{alg:DDQN}
    \begin{algorithmic}[1]
    \label{algorithm_boltzman_explore}
    \REQUIRE $\theta$ (Weights), $\alpha$ (Temperature)
    \STATE $ \Bar{\theta} \gets \theta$
    \FOR{each environment step}
        \STATE $a_t \sim \beta_\alpha(a|s_t)$
        \STATE $s_{t+1} \sim p(s_{t+1}|s_t, a_t)$ 
        \STATE $\mathcal{D} \gets \mathcal{D}\cup \{(s_t, a_t, s_{t+1})\}$ \%Append experience to buffer
    \ENDFOR
    \FOR{each iteration}
        \STATE $(s_i, a_i, s_i') \sim \mathcal{D}$ \%Sample from the experience buffer
        \STATE Compute importance weights $w$ and estimate $J_Q$, according to (\ref{eq:JQ})
        \STATE $\theta \gets \theta - \eta\nabla J_Q$  \%Gradient update
        \STATE $\Bar{\theta} \gets (1-\tau)\Bar{\theta} + \tau\theta$
    \ENDFOR
    \RETURN $\theta$
    \end{algorithmic}
\end{algorithm}
\begin{figure*}[tp]
\begin{subfigure}{.5\textwidth}
  \centering
  \includegraphics[scale=0.6]{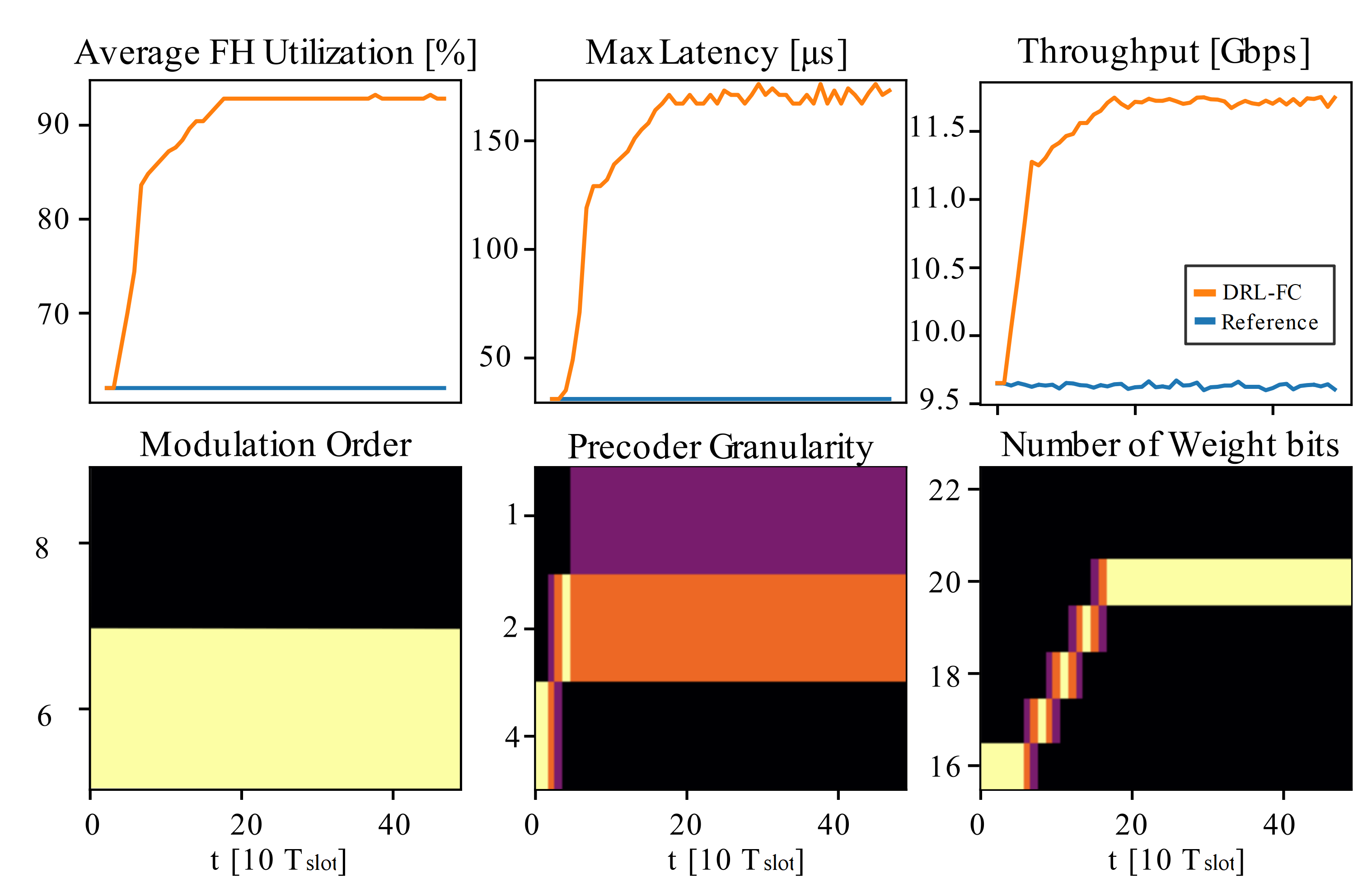}
  \caption{Medium load, $\overline{N}^{\textsc{PRB}}=175$.}
  \label{fig:sub1}
\end{subfigure}%
\begin{subfigure}{.5\textwidth}
  \centering
  \includegraphics[scale=0.6]{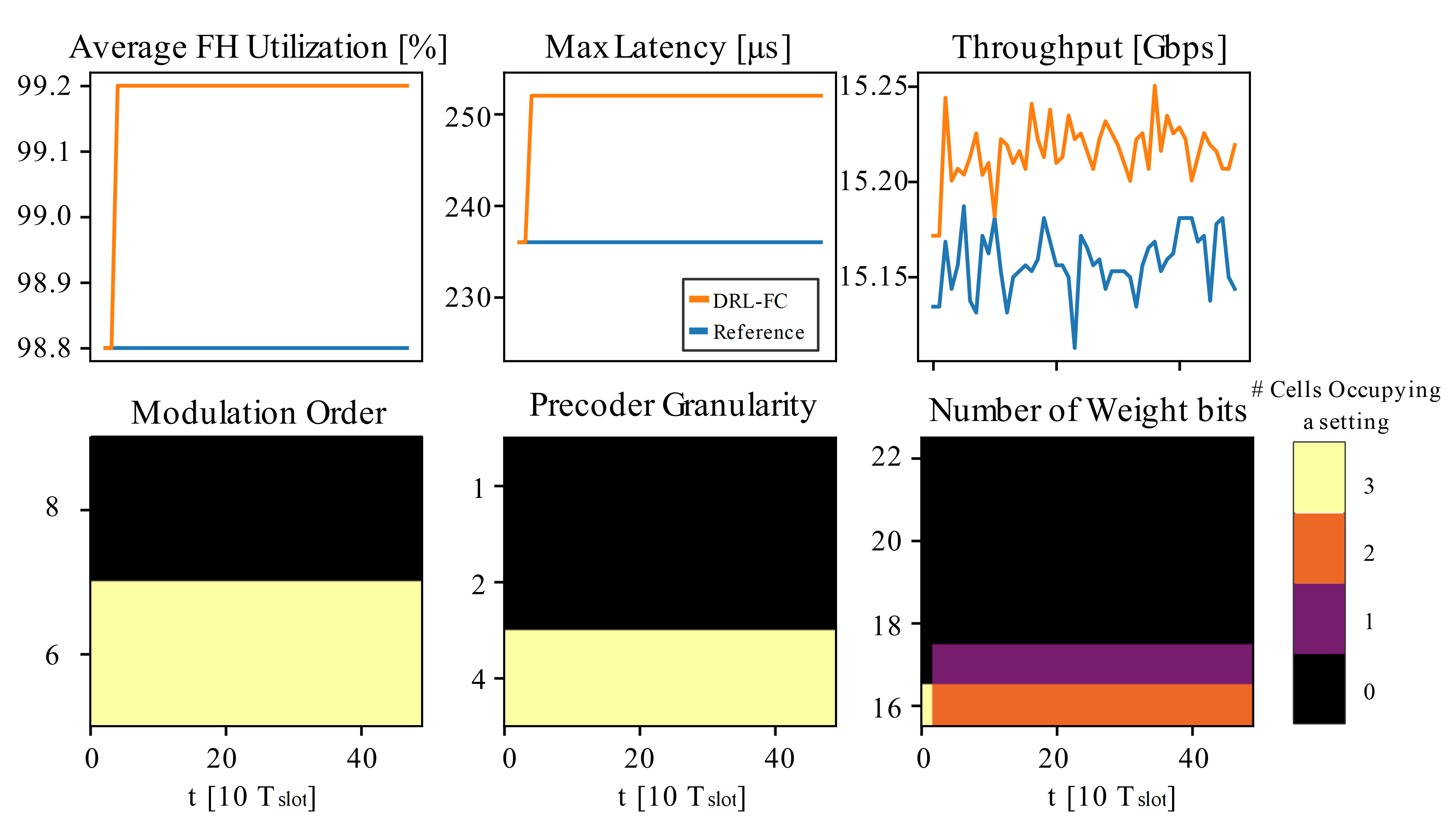}
  \caption{High load, $\overline{N}^{\textsc{PRB}}=273$.}
  \label{fig:sub2}
\end{subfigure}
\caption{Behaviour of DRL-FC and comparison to the reference scheme for (a) medium load and (b) high load. The heat maps show how many cells use a certain compression configuration. DRL-FC converges to a fixed configuration.}
\label{fig:behaviour}
\end{figure*}

\section{Numerical Evaluation}
\label{sec:numerical_evaluation}

\noindent We conduct system-level simulations to evaluate the scenario explained in Section \ref{sec:Scenario_Desc}. For this, an in-house developed Baseb{A}nd {S}ystem {S}imulator (BASS) based on ns-3 has been used which was run on a 40-core 2.8 GHz Intel(R) Xeon(R) CPU ES-2680 with 250 GB of RAM memory. BASS models the DL transmission of data and precoding weights between the BBH and the BBL via a FH based on the eCPRI protocol. BASS receives configuration messages from the AI controller and can provide various key performance indicators (KPIs) to it, e.g., air interface throughput, FH utilization, and FH latency. In this way, BASS acts as the environment for the RL agent that resides in the AI controller. The new state reflecting the impact of the taken action is signalled from BASS to the AI controller via the measured KPIs. The system parameters and values used for the simulations are shown in Table \ref{tab:Sim_Param}. The number of scheduled PRBs ($N_{t,k}^{\textsc{prb}}$) is randomized at each slot and for different cells according to a truncated Gaussian distribution, i.e., $N_{t,k}^{\textsc{prb}}\sim \min(N^{\textsc{prb}}_{\textsc{B},\mu}, \mathcal{N}(\overline{N}^{\textsc{prb}},\sigma_{N^{\textsc{prb}}}))$. Here, $\overline{N}^{\textsc{prb}}$ and $\sigma_{N^{\textsc{prb}}}$ denote the mean and the variance of scheduled PRB number, respectively. For the compression configuration $c_{t,k}$ different values can be chosen by the DRL-FC scheme according to the allowed parameter sets $\mathcal{Q}$, $\mathcal{B}^w$, along with $\mathcal{R}^w$ defined in Table \ref{tab:Sim_Param}. The DRL-FC scheme will select the most suitable compression configuration parameters to maximize the FH utilization while meeting the specified FH latency constraint. We will compare the DRL-FC scheme performance against a reference scheme in which a static compression configuration is applied to guarantee FH capacity is below $C_{\textsc{fh}}=25$ Gb/s under full load conditions. Python's Py{T}orch library has been used for automatic differentiation. 
\begin{table}
\begin{center}
\caption{Simulation parameters.}
\begin{tabular}{ |c|c|c| }
 \hline
 Parameter & Symbol & Value \\ 
 \hline
 Bandwidth & $B$ & 100 MHz\\
 \hline
 Number of available PRBs & $N^{\textsc{prb}}_{\textsc{B},\mu}$ & 273\\
 \hline
 Number of scheduled PRBs & $N_{t,k}^{\textsc{prb}}$ & 1\ldots273 \\
 \hline
 variance of scheduled PRBs & $\sigma_{N^{\textsc{prb}}}$ & 1 \\
 \hline
 Number of REs per RB & $N_{\textsc{re}}$ & $12\times14=168$\\
 \hline
 Subcarrier spacing index & $\mu$ & 1 ($\Delta f_{\textsc{scs}}=$30kHz)\\
 \hline
 Symbol duration & $T_{symb}^\mu$ & 33.33 $\mu$s\\
 \hline
 Slot duration & $T_{slot}^\mu$ & 0.5 ms\\
 \hline
 Number of cells & $K$ & 3\\
 \hline
 Number of antennas & $N_{ant}$ & 64\\    
 \hline
 Number of layers & $\upsilon_{lay}$ & 12\\ 
 \hline
 Modulation order & $q_{t,k}$ & $\mathcal{Q} =\{6,8\}$ \\    
 \hline
 Number of weight bits & $b_{t,k}^w$ & $\mathcal{B}^w =\{16,\ldots,22\}$\\    
 \hline
 Precoder granularity & $r_{t,k}^w$ & $\mathcal{R}^w =\{1,2,4\}$\\    
 \hline
 Max. latency safety parameter & $\delta$  & $10^{-3}$\\
 \hline
 Max allowed latency & $\tau_{max}$  & 260 $\mu s$\\
 \hline
 FH capacity & $C_{\textsc{fh}}$ & 25 Gb/s\\    
 \hline
 Discount factor & $\gamma$ & 0.95\\
 \hline
 Q learning rate & $\eta$ & $10^{-3}$\\
 \hline

 Soft update frequency & $\kappa$ & $5\cdot10^{-3}$\\
 \hline
 Number of parameters in NN &  &$1.8\cdot10^{6}$\\
 \hline
\end{tabular}
\label{tab:Sim_Param}
\end{center}
\end{table}

\begin{figure}
    \centering
    \includegraphics{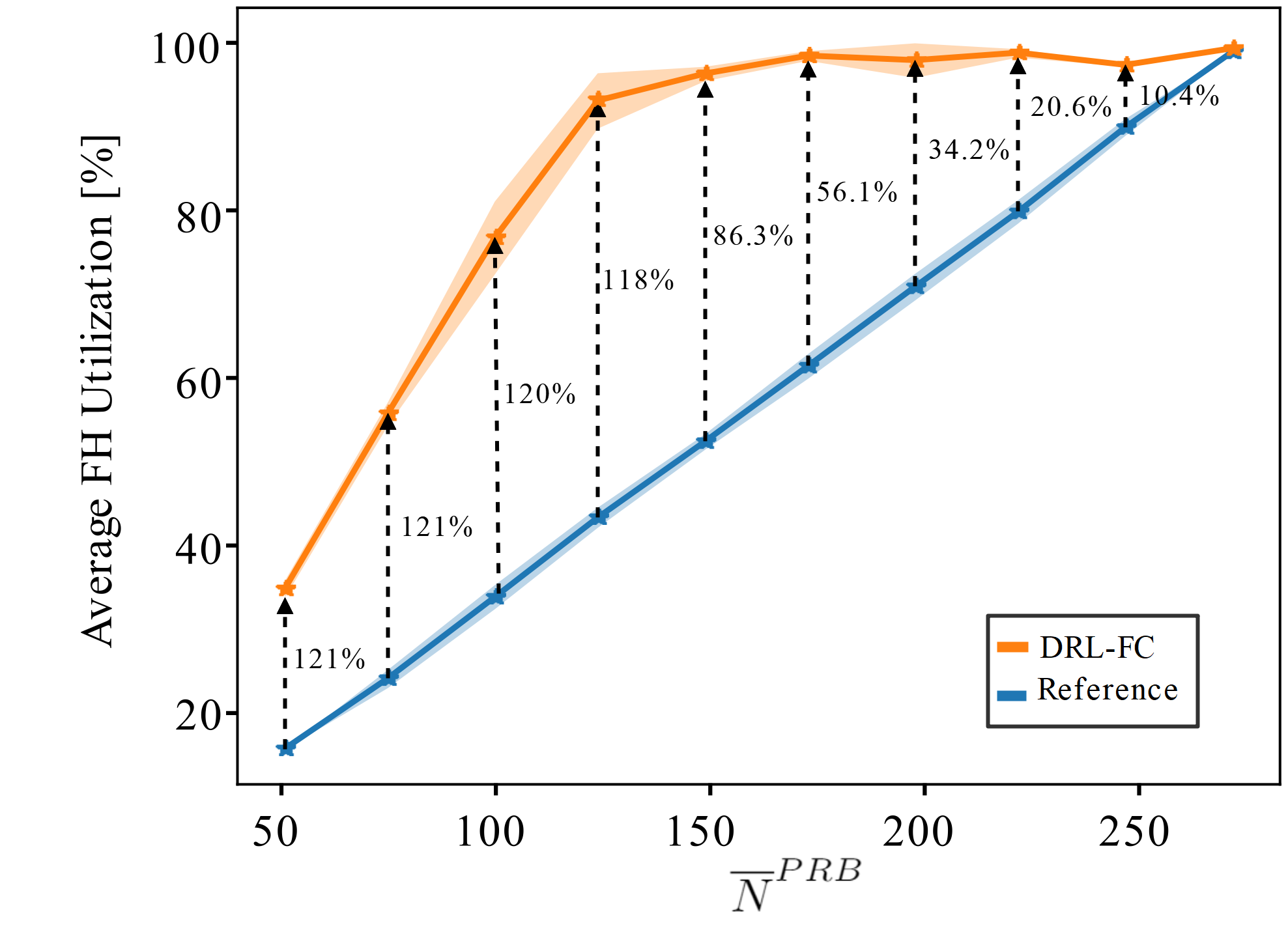}
    \caption{Average FH utilization for DRL-FC and reference scheme against mean PRB number ($\overline{N}^{\textsc{PRB}}$). The shaded regions show two sample standard deviations. The black dashed arrows show the gain in comparison to the reference policy, the annotations are the percentage gain compared to the reference policy.}
    \label{fig:gain}
\end{figure}

Fig. \ref{fig:behaviour} shows the behaviour of the DRL-FC scheme for medium and high load. Specifically, top-most subplots show the main KPIs provided by BASS against the simulation time, namely, the average FH utilization, FH latency and air interface throughput. Bottom-most subplots, on the other hand, illustrate, via a heat map, the number of cells with a certain compression configuration (i.e. modulation order, precoder granularity, and precoder weight bitwidth) as per the utilized DRL-FC scheme. For medium load, see Fig. \ref{fig:sub1}, the modulation order is set to $q_{t,k}=6$ for all cells, while precoder granularity starts with $r_{t,k}^w=4$ for all cells and then cell-wise moves to lower granularity values, changing first from $r_{t,k}^w=4$ to $r_{t,k}^w=2$ and then from $r_{t,k}^w=2$ to $r_{t,k}^w=1$ until reaching a steady configuration with $r_{t,1}^w=r_{t,2}^w=2$ and $r_{t,3}^w=1$ after $t$ exceeds roughly 20 timesteps ($\times 10$ $T_{slot}$). Similar behaviour can be observed also for the weight bitwidth, where initially all cells use $b_{t,k}^w=16$ bits and then cells one by one move to higher bitwidth values until a stable configuration is reached, i.e., $b_{t,k}^w=20$ bits for all cells. As observed in Fig. \ref{fig:sub1}, moving to lower precoder granularity and  increasing the precoder weight bitwidth leads to higher FH utilization (top-left subplot of Fig. \ref{fig:sub1}) and higher air interface throughput (top-right subplot of Fig. \ref{fig:sub1}) but also FH latency increases (top-center of Fig. \ref{fig:sub1}). Fig. \ref{fig:sub2} shows the high load case, where it can be seen that using $q_{t,k}^w=6$ and $r_{t,k}^w=4$ for three cells as well as $b_{t,k}^w=16$ bits for two cells and $b_{t,k}^w=17$ bits for one cell maximizes the FH utilization (see top-left subplot of Fig. \ref{fig:sub2}) while satisfying maximum FH latency constraint (top-centre subplot of Fig. \ref{fig:sub2}). For both medium and high load DRL-FC is compared to a reference scheme, which is dimensioned for the maximum FH load, i.e., for the case that all 273 RBs are scheduled and ensures that FH rate does not exceed the FH capacity $C_{\textsc{fh}}$ = 25 Gb/s. Clearly, DRL-FC outperforms the reference scheme both in terms of average FH utilization and air interface throughput while satisfying the maximum latency constraint.

Fig. \ref{fig:gain} depicts the average FH utilization against the average number of occupied PRB for both the DRL-FC and the reference scheme. As expected, with increasing load, i.e., number of scheduled PRBs, the FH utilization increases for both the proposed solution framework and the reference scheme until it reaches maximum for 273 PRBs. DRL-FC exhibits significantly higher FH utilization compared to the reference scheme over the wide range of scheduled PRB values due to better adaptability to varying FH loads. In particular, the highest observed FH utilization gain is obtained for a mean PRB number of $\overline{N}^{\textsc{prb}}=50$ PRBs, with a gain of 121\%. By examining Fig. \ref{fig:gain} we extract that the average FH utilization is improved 68.7\% on average, when using DRL-FC with respect to the reference case. As expected, when approaching high PRB loads, the FH utilization of DRL-FC converges to reference scheme. 

\section{Conclusions}
\label{sec:conclusions}
In this paper, we have investigated an adaptive FH compression scheme which can operate under latency constraints and limited FH capacity. We have formulated this problem mathematically as a constrained optimization problem, aimed at maximizing the FH utilization constrained on FH latency through controlling the modulation order, precoder granularity, and precoder weight bitwidth. Finding the exact solution is hard and computationally expensive as it is a combinatorial problem and requires accurate modelling, which is very difficult to obtain for realistic scenarios. Then, we have proposed a DRL-FC scheme as a data-driven solution framework, which considers the underlying system as a black-box and requires no model of the environment. The simulation results have shown that DRL-FC successfully learns a FH compression policy that maximizes FH utilization, and consequently air interface throughput, while satisfying some FH latency constraint independent of FH load. The proposed solution framework outperforms a reference scheme both in terms of FH utilization and air interface throughput. On average, the FH utilization is improved by 68.7\%. As a future work, we aim to expand the action space to account for uneven cell load scenarios and per-user compression configurations. This will require a policy gradient approach due to the increased size of action space. In addition, validating our framework over a real testbed will be also prioritized as our next steps.

\bibliographystyle{IEEEtran}
\bibliography{ref}{}

\end{document}